\documentclass[12pt]{article}
\usepackage{epsfig,amssymb}
\begin{document} 
 
\title{Sparse random matrices:\\ the eigenvalue spectrum revisited} 
\vskip 10pt 
\author{ 
Guilhem Semerjian$^{1}$ and Leticia F. Cugliandolo$^{1,2}$ 
\\ 
$^1$Laboratoire de Physique Th{\'e}orique de l'{\'E}cole Normale  
Sup{\'e}rieure,  
\\ 
24 rue Lhomond, 75231 Paris Cedex 05, France 
\\ 
$^2$Laboratoire de Physique Th{\'e}orique  et Hautes {\'E}nergies, Jussieu, \\  
1er {\'e}tage,  Tour 16, 4 Place Jussieu, 75252 Paris Cedex 05, France  \\
{\tt guilhem@lpt.ens.fr, leticia@lpt.ens.fr}
} 
\maketitle 

\begin{abstract} 
We revisit the derivation of the density of 
states of sparse random matrices. 
We derive a recursion relation that allows one to 
compute the spectrum of the matrix of incidence for finite
trees that determines completely the low concentration limit.
Using the iterative scheme introduced by Biroli and Monasson 
[J. Phys. A 32, L255 (1999)] we find an approximate expression 
for the density of states expected to hold exactly in the opposite 
limit of large but finite concentration. 
The combination of the two methods
yields a very simple geometric interpretation of
the tails of the spectrum. We test the analytic results with numerical 
simulations and we suggest an indirect numerical method to 
explore the tails of the spectrum.

\vspace{.5cm} 
\noindent LPT-ENS 02/16, LPTHE 02/10
\end{abstract} 
 
\section{Introduction}
\label{introduction}

The list of physical applications of random matrix theory is very
long. In nuclear physics, the Hamiltonians of heavy 
nuclei are replaced by large random matrices and their eigenvalues 
are associated to the energy levels. In condensed matter theory, random
matrices are used to mimic the Hamiltonians of dirty conductors
in the study of the metal-insulator transition. 
Random matrices
represent the interactions in disordered magnets as 
spin-glasses. These physical systems are intimately related to 
optimization problems and the theory of random matrices also plays an
important role in computer science. Random matrices are
the incidence matrices of random graphs and have been extensively 
studied in the context of graph theory. Consequently, the
development of 
random matrix theory and, in particular, the 
study of the spectrum of random matrices has a long 
history~\cite{Mehta,german}.

The most commonly studied random matrices are
symmetric with independent identically distributed
real elements. If the entries are 
taken from a Gaussian distribution the 
averaged spectrum of infinite matrices 
obeys the Wigner semi-circle law~\cite{Mehta,german,Edjo,efetov}. 
For such matrices studied in the thermodynamic limit, the only way to
modify this result is to 
consider the sparse case in which 
the mean number of non-zero elements on each row
remains finite when the size of the matrix diverges~\cite{Mify}. 
If the elements in the random matrix represent 
the exchanges of a model with two-body 
interactions, the fully connected case corresponds to a
matrix with non-vanishing entries outside the diagonal~\cite{SK} while 
the sparse limit corresponds to a model with finite connectivity~\cite{Vibr}.  
The latter problem is much harder than the former.
Even if replica~\cite{Robr,Robr2} and
super-symmetric~\cite{Mify,Rode} methods have been used to compute the spectrum
of dilute matrices, the derivation of a full analytic expression for it
seems out of reach. A major difference between the semi-circle law and
the density of states of dilute matrices lies in the
presence of unbounded tails in the latter case~\cite{Robr,Bauer}, 
which are absent in 
the thermodynamic limit of the former~\cite{Mehta}.

In this article we study the density of states of a sparse random
matrix using a variety of techniques. We take special advantage of 
one recently proposed by Biroli
and Monasson~\cite{Bimo}. Even if not all our results are new,
we believe that it is useful to derive them with a
method that emphasizes the underlying geometric origin
of the tails in the spectrum, namely, the fluctuations in the site 
connectivities.
In a first part, we present the results stemming from cluster
expansions~\cite{k-sat}, recovering results  mentioned by Mirlin and
Fyodorov~\cite{Mify} and Bauer and Golinelli~\cite{Bauer},
and showing how the density of states of any finite random matrix can 
be obtained in an iterative way (Section~\ref{clusters}). 
Second, in Section~\ref{replicas} we use the replica approach, following the
scheme of approximation that is based on the variational method 
developed in \cite{Bimo}. 
By emphasizing the links between the cluster expansion and the
replica functional variational method, we 
give a very simple explanation of the asymptotic form of the 
tails of the distribution derived by Rodgers and Bray~\cite{Robr}, and
we give the first steps of its more precise investigation. 
In Section \ref{numerics} we present the spectra obtained 
from the numerical diagonalisation of finite matrices. In particular, 
we demonstrate the importance of finite size effects as far as 
the dependence on the ferro or anti-ferromagnetic character of the 
interactions is concerned.
In the conclusions we summarize our results and we 
show how the study of the dynamics of dilute disordered spin models
are related to them~\cite{Secu}. 

\section{Definitions}
\label{definitions}

Let us consider an $N \times  N$ symmetric matrix with real elements
$J_{ij}$. We constrain the diagonal entries to vanish 
$J_{ii}=0$ and we denote $\lambda_k$ its 
 $N$ eigenvalues.
The elements $J_{ij}$ ($i<j$) are independent,
identically distributed random variables with a distribution 
law $P(J_{ij})$. We adopt square brackets to indicate an 
 average over this distribution,
$[\; \bullet \; ]$. The central quantity under 
study is the average density of states
\begin{equation}
\rho(\mu  ) \equiv \left[ \frac{1}{N} \sum_{k=1}^N \delta(\mu  -\mu  _k)\right] 
\end{equation}
in the thermodynamic limit, $N \to \infty$.

For most possible choices of $P(J_{ij})$
$\rho(\mu  )$ is given by Wigner's semi-circle law
except when one considers very sparse random matrices with a 
finite number of non zero elements per row in the
thermodynamic limit~\cite{Mify}. In this case,
\begin{equation}
P(J_{ij}) = \left(1-
\frac{p}{N} \right) \delta(J_{ij}) + \frac{p}{N} \pi(J_{ij})
\; ,
\end{equation}
with $p$ finite. $\pi(J_{ij})$ is a normalized 
distribution with 
average and mean square deviation of
order one (this ensures a  sensible thermodynamic limit) which 
does not contain a delta peak around zero.
$p$ is the mean number of non zero elements per
row, {\it i.e.} the mean connectivity of a given site.

In the following, we 
concentrate on a bimodal distribution
\begin{equation}
\pi(J_{ij}) = a\, \delta(J_{ij}-J_0) +
(1-a)\, \delta(J_{ij}+J_0)
\; 
\label{disorder}
\end{equation}
for the non-vanishing entries in the random matrix.
The parameter $a$ controls the asymmetry between the ``ferromagnetic'' or 
``anti-ferromagnetic'' tendency in the model. 
Since we can absorb the dependence on $J_0$ 
with a global rescaling of the
density of states, 
$\rho(\mu  ) = (1/J_0) \tilde{\rho}(\mu  / J_0)$, we set
$J_0=1$ henceforth.

\section{Cluster expansion}
\label{clusters}

Given a particular realization of the matrix $J$ among the random
ensemble, one can associate it to a graph in a very natural
way. Consider $N$ sites labeled from $1$ to $N$ and draw a link between
the sites $i$ and $j$ if the entry $J_{ij}$ is non zero. The value $J_{ij}$
of the interaction can be written
next to the link. Two sites are called {\it adjacent} if there is
a link between them, and {\it connected} if there is a path of
adjacent sites from one to the other. A {\it cluster} is a set of
connected sites, disconnected from all others. 

These definitions
make the relation between random matrices and random
graphs explicit. As a matter of fact,  
the study of the 
ensemble of such graphs (forgetting the values of the interactions) is
nothing but the
well known random graph problem.
When $p<1$ the random matrix $J$ is so 
sparse that all sites belong to finite clusters. At $p=1$ a 
 percolation transition occurs and a giant cluster containing a 
finite fraction of the sites  appears. The 
rest of the sites belong to finite clusters~\cite{percolation}. 

In this Section 
we derive a cluster expansion of the density of states, {\it
i.e.} a development in powers of the mean connectivity $p$ of
$\rho(\mu  )$. This expansion is not valid beyond the percolation
threshold; still, some results derived here give a geometrical insight into the
behavior of $\rho$ for all $p$, at least in the limit $|\mu |  \to \infty$.

If one orders the sites by grouping those belonging 
to the same cluster, the matrix $J$ acquires a block-diagonal form,
each cluster being associated to one block.
The density of eigenvalues for a particular realization of the
ensemble, $\rho_J(\mu  )$, can then be written as a double sum,
\begin{equation}
N \rho_J(\mu  ) = \sum_{{\cal C}} \sum_{k_{\cal C}}
\delta(\mu    -\mu  _{k_{\cal C}}) 
\; ,
\label{sum}
\end{equation}
where the index ${\cal C}$ labels the clusters and $k_{\cal C}$ labels
the eigenvalues of each cluster.
Under this form, $\rho_J(\mu  )$ is  
an additive quantity over clusters and we can apply the 
technique described in \cite{k-sat} to transform this expression 
into a simpler one. Indeed, clusters can be separated into 
ensembles with the same topology, 
ignoring for the moment the assignment of the values of the interactions. 
The averaging over different realizations of the random matrix
proceeds in two steps; one
first chooses the topology of the cluster, with its associated
probability, and, subsequently, one averages over the values 
of the interactions with the distribution $\pi(J_{ij})$. 
For a given cluster, once the
latter average is performed, the density of states 
depends only on its topology.
 (Note that one could do so for any distribution $\pi$,
even if not bimodal.) This remark allows us to rewrite the average of the
sum in  Eq.~(\ref{sum}) in a more convenient manner.
If we introduce an index $t$
that labels all possible topologies,
$n_t(J)$ the number of $t$-like clusters present in a given
realization of the ensemble,  
and $\overline \rho_t$ 
the average over the distribution $\pi$ of 
the density of states of the $t$-like cluster $\rho_t$,
we obtain the following
expression for the averaged density of eigenvalues:
\begin{equation}
\rho(\mu  )=
[\rho_J(\mu    )]= \sum_t \, \frac{[n_t]}{N} \, L_t \, \overline \rho_t(\mu    )
\; .
\end{equation}
We included the isolated sites associating them
to the index $t=0$ and we denoted $[n_t]$ the
average number of $t$-like clusters.
As in \cite{k-sat} 
we introduce a function $X_t^i$ that takes the value
$1$ if the site $i$ belongs to
a $t$-like cluster and $0$ otherwise and we call  
$L_t$ the number of sites in a $t$-like cluster. Then
\begin{equation}
n_t(J)=\frac{1}{L_t} \sum_i X_t^i
\; ,\;\;\;\;\;\;\; \Rightarrow
\;\;\;\;\;\;\;\;
\frac{1}{N} \, [n_t] = 
\frac{1}{L_t} [X_t^1] = \frac{P_t}{L_t}
\; ,
\end{equation}
where $P_t\equiv [X_t^1]$ is the probability that a given variable
belong to a $t$-like cluster. Finally,
\begin{equation}
\rho(\mu  )  = 
\sum_t P_t \; \overline \rho_t(\mu  ) =
\sum_t \frac{P_t}{L_t}  \; 
\overline{ \sum_{l=1}^{L_t} \delta (\mu    -\mu  _l) }
\; ,
\end{equation}
where $\mu  _l$ are the eigenvalues of the $t$-like cluster, $P_t$ is given by
$P_t = p^{L_t-1} e^{-p L_t} K_t$,
and $K_t$ is a symmetry factor~\cite{k-sat}. 
The overline denotes an average over the distribution $\pi$, for
fixed topology.
Note that only tree-like
clusters contribute to the thermodynamic limit: the probability that a
site belong to a cluster containing a loop of finite length is of
order $1/N$. Reordering this series in powers of $p$, it is clear that
all clusters of size $L_t \le m+1$ have to be considered to obtain the
expansion at order $p^m$.

For the bimodal $\pi(J_{ij})$ each type of cluster
contributes a sum of a finite number of delta functions to 
$\rho(\mu  )$.
The location of the delta functions are symmetrically distributed
around $\mu  =0$. The only task remaining in order to obtain the series
expansion is to diagonalize the finite size matrices
corresponding to each cluster. For
instance, up to first order in $p$, one finds (for any value of $a$):
\begin{equation}
\rho(\mu  ) = \delta(\mu    ) + \frac{p}{2} 
\left[\delta(\mu    -1) - 2 \delta(\mu    )  + \delta(\mu    +1)\right] 
\; .
\end{equation}

There does not seem to be a straightforward way to compute the
eigenvalues of an arbitrarily shaped tree-like cluster. A possible 
method would be to solve the integral equation arising from the 
replica or the super-symmetric approach perturbatively
and to expand the order parameter
in powers of the concentration, as explained in~\cite{Mify}. Another
method, exposed below, is to establish recursion relations on the
characteristic polynomial of finite size matrices~\cite{DeRo}.

\begin{figure}
\centerline{
\epsfig{file=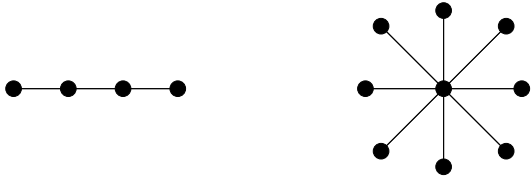,width=7cm}}
\label{sketch}
\caption{Two examples of finite clusters: 
a linear ($n=4$) and a star-like one ($k=8$).}
\label{examples}
\end{figure}

Before getting into the general derivation, 
let us study two examples of simple clusters for which one can
compute the eigenvalues explicitly.
Consider first a linear cluster with $n$ sites (see the left panel of
Fig.~\ref{examples}). For this structure
the corresponding block in the matrix $J$ has
elements $\pm  1$ on the two diagonals next to the main one, and $0$
everywhere else. 
It is then a simple exercise of linear algebra to show that the $n$
eigenvalues are
\begin{equation}
\mu  _l= 2 \; \cos \left(\frac{l \pi}{n+1}\right) \; , 
\;\;\;\;\: \: \: l = 1,...,n
\; .
\end{equation}
Note that for any value of $n$, the eigenvalues belong to $[-2,2]$.
The ``opposite'' kind of geometry is given by clusters with one site connected
to $k$ neighbours (see the right panel in Fig~\ref{examples}). 
Again the matrix $J$ is readily written. 
One easily shows that the spectrum of such a cluster has $k-1$
vanishing eigenvalues, one eigenvalue given by $\sqrt{k}$ and another one
given by $-\sqrt{k}$.
This behavior is very different from the one of the linear
clusters: more and more branched clusters yield larger and larger
eigenvalues. Very important for the arguments developed in
Section~\ref{replicas} is the
fact that the eigenvectors associated with
the eigenvalues $\pm  \sqrt{k}$ have a larger magnitude on the central 
site than on its neighbours, with a ratio $\sqrt{k}$.
Thus, for large
$k$, the eigenvectors are more and more concentrated on the central
site. We shall come back to this point later, when discussing the
behavior of the density of states at large eigenvalues.

\begin{figure}
\centerline{
\input{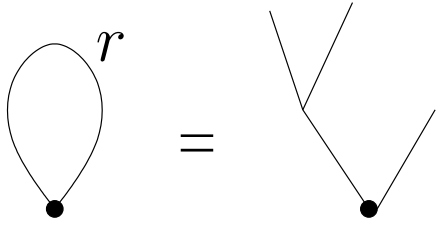}}
\caption{An example of rooted tree}
\label{ex_root}
\end{figure}

\begin{figure}
\centerline{
\input{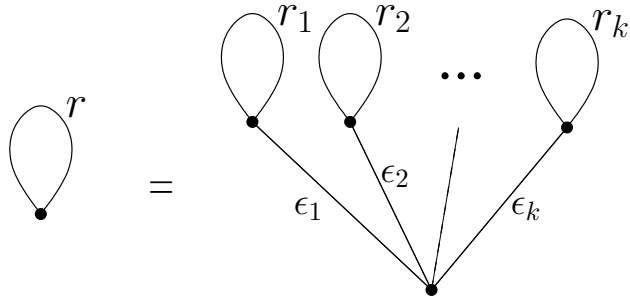}}
\caption{Decomposition of a rooted tree}
\label{root2}
\end{figure}

To derive the general recursion relation we focus on  
tree-like structures since loops of finite size disappear in the
thermodynamic limit. We define a {\it rooted tree} $r$
as a tree with one of its sites (the root) particularized. We
represent a generic rooted tree as a bubble englobing the tree, 
with only the root shown (see Fig.~\ref{ex_root}).
If the root of the tree $r$ is connected to, say,  $k$ other sites that we 
call its neighbours, we denote $r_i$,
$i=1,...,k$, the trees whose roots are the neighbours of the original root
and $\epsilon_i$ the values of the edges that link the 
original root to the $k$ secondary ones (see Fig.~\ref{root2}).

We call $P_r(\lambda)$ the characteristic polynomial of the
matrix associated with the rooted tree $r$, and $\tilde P_r(\lambda)$
the one of the matrix in blocks associated to the disconnected trees
obtained after deleting the root of $r$ and the edges emerging from it. 
We thus have
\begin{equation}
\tilde P_r(\lambda)=\prod_{i=1}^k P_{r_i}(\lambda)
\; .
\end{equation}
With these definitions the recursion relation is obtained by expanding
the determinant along the rows and the columns of the roots, and reads
\begin{equation}
P_r(\lambda)= - \lambda \prod_{i=1}^k P_{r_i} (\lambda) - \sum_{i=1}^k
\epsilon_i^2 \tilde{P}_{r_i} (\lambda) \prod_{j \ne i} P_{r_j} (\lambda)
\; .
\label{recursion}
\end{equation}
Our first remark is that the parameters $\epsilon_i$ appear only
through their squares, so we can prove by recurrence that for any
tree, the average spectrum for the bimodal distribution (\ref{disorder}) 
does not depend on the value of $a$. Thus, when $p<1$ and the 
thermodynamic limit is taken $\rho(\mu  )$ 
is totally independent of the ferromagnetic or anti-ferromagnetic 
character of the distribution.
When $p>1$, loops of length of order $\ln N$ appear in
the giant cluster and the argument above, relying on the tree
structure of the interactions, is not valid anymore. We shall develop
this point in Section~\ref{nonsym}. 

\begin{figure}[hb]
\centerline{
\input{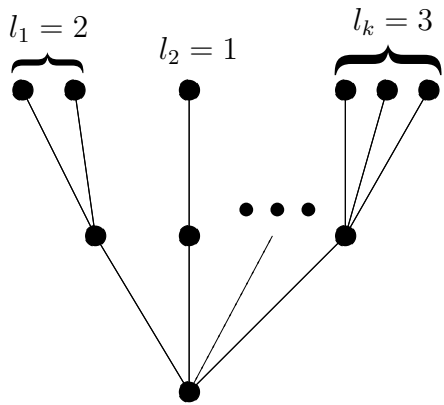}}
\caption{Two level tree}
\label{tltree}
\end{figure}

As a simple example, 
let us now write down the characteristic polynomial for the two level
tree sketched in Fig.~\ref{tltree}. The root has 
$k$ neighbours and  each of them has $l_i+1$ neighbours, $i=1,...,k$.
If we define $L\equiv\sum_i l_i$, the characteristic polynomial is given by

\begin{equation}
P(\lambda)=(-\lambda)^{L-k+1} \left[ \prod_i (\lambda^2-l_i) - \sum_i
\prod_{j \ne i} (\lambda^2-l_j) \right]
\label{eq_tltree}
\end{equation}
whose roots give the $L+k+1$ eigenvalues of the corresponding matrix.
For instance in the symmetric case ($l_i=l \: \: \forall
i$), the eigenvalues (and their degeneracies) are: $0$ ($k(l-1)+1$) ,
$\pm  \sqrt{l}$ ($k-1$ for each sign) and $\pm  \sqrt{k+l}$ ($1$ for
each sign). 

In principle, the relation (\ref{recursion}) can be iterated to obtain
the characteristic polynomial of any tree, and from it its eigenvalues.

To conclude this Section, let us note than even for $p>1$, a finite
fraction of sites belong to finite-size clusters. Thus, 
the spectra contains delta peaks located at eigenvalues of finite tree
clusters for all $p$, as established
in~\cite{Bauer,Kieg} 
(it is also shown in \cite{Bauer} that parts of the giant
cluster contribute to these delta peaks). 

In the context of the gelation transition, a detailed study of the low
concentration phase for related matrices was performed in~\cite{Zip}.

\section{Replica approach}
\label{replicas}

In this Section we compute the spectrum using the replica method. We
first give a qualitative argument to suggest where one can expect to find
localized states. Then we describe the method and find a series of
approximate expressions for $\rho(\mu)$.

\subsection{Localized eigenvectors on the Bethe lattice}
\label{qualitative}

One cannot use the above cluster expansion
to derive the density of states above the percolation 
transition $p=1$ where a 
giant component with a finite fraction of the sites appears in
the thermodynamic limit. 
The giant cluster contains loops with 
length  $O(\ln N)$ so, 
locally, it has a tree-like structure~\cite{percolation}. 

It is interesting to know if some eigenvectors of the sparse matrix
are localized, {\it i.e.} if they have a finite fraction of their
weight on a finite number of sites. A qualitative answer to this
question can be given by assuming that one site has $k$ 
neighbours,
and that all the other sites have the same connectivity $p+1$ (the
average connectivity of a site having already one neighboor is $p+1$),
with no loops. That is to say, we approximate the giant component 
with a Bethe lattice with a
single ``defect'', the site with a connectivity that is different from the
mean. It is then easy to derive a condition for the existence of an 
eigenvector localized on this defect and with a spherical symmetry
{\it i.e.} such that its magnitude on each site only depends on the distance
to the central site.
For the
ensemble of matrices under consideration, one finds the condition:
$k \geq 2p$. Thus, we expect localized eigenstates to appear around sites
with connectivities much larger than the mean. 
Note that for the Laplacian of a random graph (instead of the incidence matrix
considered here) the same argument leads to $|k-p| \geq \sqrt{p}$~\cite{Bimo}
for the localization
condition of eigenstates; the required 
deviation of the local connectivity from the mean  is less stringent
in this case. 

This argument can be refined by considering a site with $k$
neighbours, each having $l+1$ neighbours, and all other sites having
the mean connectivity $p+1$. Then the condition for the existence of a
localized eigenstate is : $k+2l \geq 4p$. Of course if one takes
$l=p$, the above condition is recovered. If $l$ is slightly lower
than the mean, the condition on $k$ becomes more and more
demanding. Conversely, if $l$ is slightly higher than the mean, the
condition on $k$ is relaxed. One can expect the latter case to be
privileged in random graphs for which the distribution of connectivities
is Poissonian: sites with connectivity of order twice the mean are
rare, but their neighbours are likely to have a connectivity slightly
above the mean.
 
\subsection{Method and definitions}

Field theories, either replicated~\cite{Edjo} or
super-symmetric~\cite{german,efetov}, have been used to determine the average
spectrum of random matrices. For sparse matrices, the result is
expressed in terms of the solution to an integral equation, which has
proven too difficult to be solved exactly. Approximate results, either
perturbative or non perturbative~\cite{Robr} 
with respect to the inverse
concentration $1/p$ have been obtained. We shall revisit this 
problem using the replica approach and the 
scheme of approximation developed by Biroli and
Monasson~\cite{Bimo} (see also~\cite{BrHu} for a similar scheme). This
method will
allow us to unveil the geometric origin of the tails in the 
distribution 
and to compare with results established above for finite size clusters.
For simplicity,  we
focus on the symmetric bimodal distribution introduced in 
(\ref{disorder}) with $a=1/2$. We come back to the dependence on $a$
in Section~\ref{nonsym}.

The normalized density of states for a particular
realization of disorder can be computed in terms of the
resolvent~\cite{Mehta},
\begin{equation}
\rho_J(\mu  ) = \frac{1}{N\pi} \mbox{Im Tr} \left[J_{ij}- (\mu    
+i \epsilon) 
\delta_{ij} \right]^{-1}
= \frac{2}{N\pi}  \left. \mbox{Im}  \frac{\partial}{\partial\mu  }
\ln Z(\mu  ) \right|_{\mu    +i\epsilon}
\; ,
\label{definition-rho}
\end{equation}
with $\epsilon>0$ and infinitesimal, and
\begin{equation}
Z(\mu  ) = \int \prod_i d\phi_i \; \exp\left( \frac{i\mu  }{2} \sum_i \phi_i^2 - 
 \frac{i}{2} \sum_{ij} J_{ij} \phi_i  \phi_j \right)
\; .
\end{equation}
The average over  disorder, $[\rho_J (\mu  )]$, 
is evaluated by
introducing replicas to average the logarithm, $\lim_{n\to 0} \ln [Z^n]
= n[\ln Z]$. A very useful way to study the mean field theory of
dilute systems~\cite{cphi} 
relies on introducing a function $c(\vec \phi)$ that measures 
the fraction of sites with a field $\vec{\phi_i}$ equal to a chosen 
value $\vec{\phi}$, the vector denoting the $n$-dimensional replica space,
\begin{equation}
c(\vec \phi) \equiv \frac1{N} \sum_i \delta(\vec \phi -\vec \phi_i) 
\; .
\end{equation}
Thus, 
\begin{equation}
\rho(\mu  ) = \lim_{n \to 0} \frac{2}{nN\pi}  \left. \mbox{Im}  \frac{\partial}{\partial\mu  }
\ln \left[Z^n\right] \right|_{\mu    +i\epsilon}
\end{equation}
with
\begin{eqnarray}
[ Z^n ] &=& \int {\cal D}c(\vec \phi) \, \exp\left( 
-N S_{\mbox{\tiny{\sc eff}}}(c(\vec\phi)) \right)
\; ,
\nonumber\\
S_{\mbox{\tiny{\sc eff}}}(c(\vec\phi)) &=&
\int d\vec \phi \, c(\vec \phi) \ln c(\vec \phi) -
\frac{i\mu  }{2} \int d\vec \phi \, c(\vec \phi) {\vec{\phi}}^2 +
H_{\mbox{\tiny{\sc eff}}}(c(\vec \phi))
\; ,
\label{pathI}
\end{eqnarray}
where the integration is taken only over normalized $c(\vec \phi)$. 
The effective Hamiltonian is defined as 
\begin{equation}
\exp\left(- N H_{\mbox{\tiny{\sc eff}}}\right)=
\left[ \exp\left(-\frac{i}{2} \sum_{ij} J_{ij} \phi_i
\phi_j \right) \right]_J
\end{equation}
and for the ensemble of symmetric sparse random matrices 
with a bimodal symmetric distribution 
it reads
\begin{equation}
- H_{\mbox{\tiny{\sc eff}}}(c(\vec \phi)) = -\frac{p}{2} + \frac{p}{2} 
\int d\vec \phi\: d \vec \psi\: c(\vec \phi)\: c(\vec \psi) 
\cos(\vec \phi \cdot    \vec \psi)
\; .
\end{equation}

In the limit of large sizes, $N \to \infty$,
the path integral in Eq.~(\ref{pathI}) can be computed with the
saddle-point method  and it is 
dominated by the neighbourhood of the extreme, $c^{\mbox{\tiny{\sc
sp}}}(\vec\phi)$,  
of the effective action $S_{\mbox{\tiny{\sc eff}}}$, which satisfies:

\begin{equation}
c(\vec\phi) = \mathcal{N} \exp \left[\frac{i}{2} \mu \, \vec{\phi}^2 -
\frac{\delta H_{\mbox{\tiny{\sc eff}}}}{\delta c(\vec\phi)}\right] \;,
\label{eq_sp}
\end{equation}

with $\mathcal{N}$ a normalization constant.
 
Then, the averaged density of states reads
\begin{equation}
\rho(\mu  ) =\lim_{n \to 0} \frac{1}{\pi n} \mbox{Re} \int d\vec\phi 
\; c^{\mbox{\tiny{\sc sp}}}(\vec\phi) \, \vec{\phi}^2
\; .
\end{equation}
Even with the assumption of invariance of $c(\vec \phi)$ with respect to
rotations in the $n$-dimensional replica 
space, it has proven too difficult to solve the saddle point
equation explicitly. A numerical solution of a similar equation was
given~\cite{INM} in the context of the study of the instantaneous
normal modes in a liquid.

\subsection{Effective medium approximation}

The first step of the approximation scheme~\cite{Bimo} consists 
in solving the problem
variationally, {\it i.e.} by restricting the original problem of 
extremization to the particular Gaussian 
subspace of the functions $c(\vec{\phi})$. 
This first step is called the ``effective medium approximation'' ({\sc ema})
since it amounts to assuming that all the sites are equivalent and 
play the same role. Thus
one replaces $c(\vec \phi)$ with a Gaussian
{\it Ansatz}
\begin{equation}
c^{\mbox{\tiny{\sc ema}}}(\vec \phi) = (2 \pi i \sigma(\mu    ))^{-n/2} \exp\left( -
\frac{\vec{\phi}^2}{2 i \sigma(\mu    )} \right)
\; ,
\end{equation}
with $\sigma(\mu    )$ the variational function determined
by the stationarity condition 
\linebreak
$\delta S_{\mbox{\tiny{\sc eff}}}/\delta \sigma(\mu    )=0$. 
After some algebra, we find that $\sigma=\sigma(\mu  )$ is given by 
the cubic equation
\begin{equation}
\sigma^3 + \frac{p-1}\mu   \sigma^2 - \sigma + \frac{1}\mu  =0
\; .
\label{cubic-eq}
\end{equation}
Among the three roots of the equation, one has to choose the
one with \linebreak
$\mbox{Im}\: \sigma(\mu  + i \epsilon) < 0 $ for the integrals to converge.
The average density of states reads 
$\rho^{\mbox{\tiny{\sc ema}}}(\mu  ) = 
-1/\pi \; \mbox{Im} \, \sigma(\mu  +i\epsilon)$.
It is easy to check that  
these equations yield the correct Wigner semi-circle law in the
fully connected limit, when $J_0\to
J_0/\sqrt{N}$ and $p\to N$. The {\sc ema} is exact in this case.

In the following we focus on the case $p>1$. For $p<1$, the sites
belong to finite size clusters and they are very heterogeneous in
nature so we do not expect the {\sc ema} to yield accurate 
results. Moreover the cluster expansion is exact in this regime
and it is enough to solve the problem completely. The approach used
here should be valid for $p$ large but finite, as the larger the value
of $p$, the smaller the fraction of sites in finite size clusters.

Solving the cubic equation, one finds that for $|\mu  |\leq\lambda_c$ 
the density of states is 
given by the continuous function 
\begin{eqnarray}
\rho^{\mbox{\tiny{\sc ema}}}(\mu  ) &=& \frac{\sqrt{3}}{2 \pi}  
\sqrt[3]{-\left(\frac{p-1}{3\mu  }\right)^3 - \frac{p+2}{6\mu  } +
\sqrt{\frac{(\lambda_c^2-\mu  ^2)(\mu  ^2+\alpha^2)}{27 \mu  ^4}}} 
\nonumber\\
&& - \frac{\sqrt{3}}{2 \pi} \sqrt[3]{-\left(\frac{p-1}{3\mu  }\right)^3 - \frac{p+2}{6\mu  } -
\sqrt{\frac{(\lambda_c^2-\mu  ^2)(\mu  ^2+\alpha^2)}{27 \mu  ^4}}}
\; ,
\label{rho-ema}
\end{eqnarray}
and vanishes outside the band
$[-\lambda_c,\lambda_c]$, with  
\begin{equation}
\lambda_c=\sqrt{\frac{-p^2 + 20 p + 8 +\sqrt{p(p+8)^3}}{8}}
\; , \: \: \:
\alpha^2 = \frac{p^2-20p-8+\sqrt{p(p+8)^3}}{8} \; ,
\end{equation}
see the solid curve in Fig.~\ref{figure_rho}. The density of states in the
{\sc ema} vanishes at $\pm  \lambda_c$ as a square root,
$\rho^{\mbox{\tiny{\sc ema}}}(\mu  ) \sim (\lambda_c - |\mu
|)^{1/2}$, just as for a
semi-circle law. Expanding Eq.~(\ref{rho-ema}) in powers of $1/p$ 
one recovers the first orders of the $1/p$ expansion of Rodgers and 
Bray~\cite{Robr}: 
\begin{eqnarray}
\lambda_c &\sim& 2 \sqrt{p} \left(1+\frac{1}{2p}\right)
\; ,
\\
\rho(\mu  ) &\sim& \frac{2}{\pi \lambda_c^2}\sqrt{\lambda_c^2-\mu  ^2}
\left[1+\frac{1}{p} \left(1-\frac{4\mu  ^2}{\lambda_c^2}\right)\right]
\; .
\end{eqnarray}
It would be interesting to check if the {\sc ema} and the 
perturbative solution of these authors 
lead to the same density of states at all
orders in the expansion in powers of $1/p$. If this were true,
the {\sc ema} would be the exact resummation 
of the perturbative solution in~\cite{Robr}. 

Note that at this level of the approximation scheme there are no
states outside the band $[\lambda_c,\lambda_c]$, which is consistent
with the qualitative argument given in Section~\ref{qualitative}:
localized states with
high eigenvalues are due to fluctuations of the local connectivity,
whereas, by definition, the {\sc ema} assumes a uniform local connectivity.

\subsection{Single defect approximation}

To go beyond the {\sc ema}, it is useful to write the saddle point
equation (\ref{eq_sp})
on $c(\vec{\phi})$ in the following form
\begin{equation}
c(\vec{\phi}) = \eta \: e^{\frac{i}{2} \mu  \vec{\phi}^2}
\sum_{k=0}^{\infty} \frac{e^{-p}p^k}{k!} \left[ \int d\vec{\psi}
\; c(\vec{\psi}) \cos( \vec{\phi} \cdot  \vec{\psi}) \right]^k
\; ,
\label{iter-sda}
\end{equation}
with $\eta$ a normalization constant tending to $1$ as $n \to 0$.
This equation has the following geometric interpretation: any chosen
site has $k$ neighbours with probability $\exp(-p) p^k / k!$ and 
each of them interact with the central site through the effective
Hamiltonian. The self consistent equation is obtained by imposing the
equality of the distribution of effective fields $c(\vec{\phi})$ for
all sites.
 
In the single defect approximation, one uses the above saddle-point
equation in an iterative form, {\it i.e} one 
inserts the result of the Gaussian approximation in the right-hand
side of Eq.~(\ref{iter-sda}). This means that one allows the
connectivity of a given site to fluctuate but its neighbours are
treated as part of the effective medium. The procedure  yields
\begin{equation}
\rho^{\mbox{\tiny{\sc sda}}}(\mu  )= \sum_{k=0}^\infty 
\frac{e^{-p} p^k}{k!} \, 
\left(-\frac{1}\pi\right) \, \mbox{Im} \frac{1}{\mu  +i\epsilon - k 
\sigma(\mu  +i\epsilon)} 
\; .
\label{rho-sda}
\end{equation} 

\begin{figure}[h]
\centerline{
\input{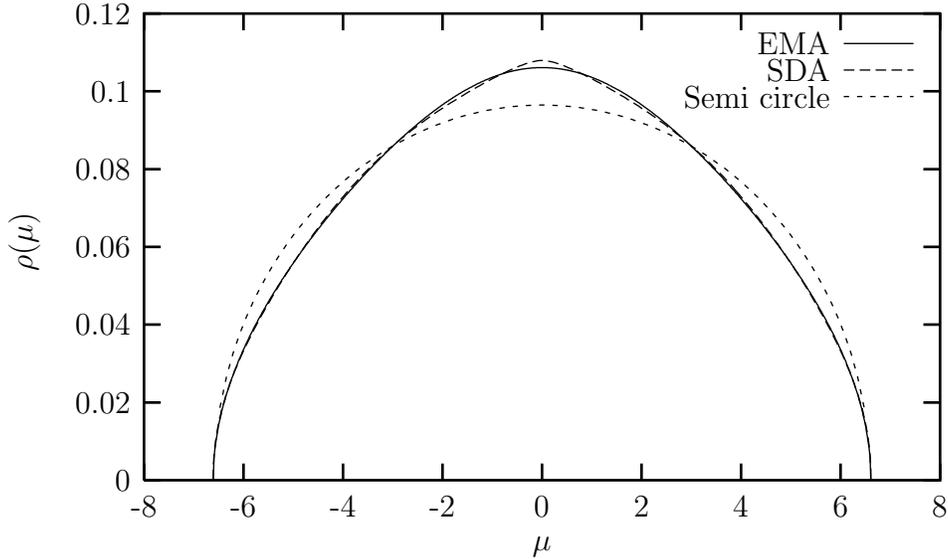}}
\caption{Extended part of the density of states for $p=10$, in the 
{\sc ema} and {\sc sda} ($\lambda_c \approx 6.60$). A semi-circle of
same width is shown for comparison.}
\label{figure_rho}
\end{figure}

For $\mu  \in [-\lambda_c,\lambda_c]$, the {\sc sda} density of states is
different from the {\sc ema} one; yet it still vanishes at $\pm 
\lambda_c$ as a square root. In Fig.~\ref{figure_rho} we show the 
central band of the density of states in the {\sc ema} and the 
{\sc sda} for $p=10$; for comparison, we include a semi-circle 
with the same support. It is clear from the figure
that the main modification introduced by the {\sc sda} is concentrated
around $\mu  \sim 0$. 

Outside the band, delta peaks appear when
the denominator of the above expression vanishes. They are
located at $\pm  \mu  _k$, with
\begin{equation}
\sigma(\mu  _k) = \frac{\mu  _k}{k} 
\; \: \: \: , \; \: \: \:  |\mu  _k| \geq \lambda_c \: \: ,
\label{eq_peaks_sda}
\end{equation}
and the weight of the peak at $\mu  =\pm  \mu  _k$ is
\begin{equation}
\frac{e^{-p} p^k}{k!} \frac{1}{1-k \sigma'(\mu  _k)}
\; .
\end{equation}

As $\sigma$ is a real uneven decreasing function of $\mu  $ for
$|\mu  |\ge \lambda_c$, there is a minimum value of $k$ for which
Eq.~(\ref{eq_peaks_sda}) has a solution.
A sufficiently large fluctuation of the local connectivity is necessary
for localization, as predicted by the qualitative argument given
in Section~\ref{qualitative}. 
For instance, for $p=10$, the first peak is found for
$k_{min}=24$. One can check that $k_{min} \sim 2 p$ when $p \to
\infty$ and the condition derived by the qualitative argument of
Section~\ref{qualitative} is recovered in this limit. 

Since the peaks extend all the way up to $|\mu  |\to\infty$
(as $k$ grows, the peaks move away from $\lambda_c$)
we  can now work out the asymptotic behaviour of the density of states.
Far from the continuous band, {\it i.e.} when $k\to\infty$ and 
consequently $|\mu  |\to\infty$, we have $\sigma(\mu  ) \sim 1/\mu  $.
The peaks are thus located at $\mu  _k \sim \pm  \sqrt{k}$
and their weight is $\exp(-p) p^k /(2 \, k!)$ for both signs.
In this limit,  the 
distance between the peaks decreases, since $(\sqrt{k}-\sqrt{k-1})
\sim 1/(2\sqrt{k}) \to 0$  and one can propose a
continuous approximation to describe the tails of the 
distribution:
\begin{equation}
\rho^{\mbox{\tiny{\sc sda}}}(\sqrt{k}) (\sqrt{k}-\sqrt{k-1}) \sim 
\frac{1}{2} \, \frac{e^{-p}p^k}{k!} 
\; .
\end{equation}
After expanding for large $k$ and changing variables $k \to
\mu  ^2$,   one obtains the asymptotic form 
\begin{equation}
\rho^{\mbox{\tiny{\sc sda}}}(\mu  ) \sim \frac{1}{\sqrt{2\pi}} \;  
e^{-p} \exp\left[ -\mu  ^2 \ln \left( \frac{\mu  ^2}{pe} \right) \right]
\;\;\;\;\;\;\;
\mu  \gg  \lambda_c
\; .
\label{tails}
\end{equation}
This result was obtained by Rodgers and Bray in~\cite{Robr}, after an
involved non perturbative treatment of an integral equation following
a method proposed in~\cite{KimHarris}.

Let us emphasize the very simple geometric interpretation of this
asymptotic form unveiled by the present derivation. A given site in a
random graph has $k$ neighbours with probability $\exp(-p) p^k
/k!$. For large $k$, there are two eigenvectors localized on it, with
eigenvalues $\pm  \sqrt{k}$. It is then not surprising that in the limit of
large $k$, one finds the same eigenvalues as for a star-like {\em finite}
cluster. When $k$ grows, the eigenvector is more and more
concentrated on the central site and the
eigenvalue is less and less sensitive to the environment of the first
neighbours; if they are disconnected from the rest of the
sites, as for finite clusters, or inside the giant cluster 
is not important in this limit~\cite{footnote}.

We would like to stress a similarity with a result
in~\cite{Bauer}. These authors express the moments of $\rho(\mu  )$ 
as polynomials in $p$ and they provide a bound for the coefficients
of these polynomials. Their lower bound is expressed in terms of 
Stirling numbers and polynomials that  are
the moments of the distribution \linebreak
$f_p(x) = \frac{e^{-p}}{2} \sum_{k=0}^{\infty}
\frac{p^k}{k!} (\delta(x-\sqrt{k}) + \delta(x+\sqrt{k}))$ that we
argued to be the asymptotic form of the eigenvalue spectrum.
Moreover, Stirling numbers also arise when considering 
star-like geometries via the enumeration of
particular walks on these trees. There should be a deep connection 
between these two approaches beyond this formal similarity. 

\subsection{Two level approximation}

Let us now briefly explain how to improve on the above results. The next 
step in the approximation scheme consists in 
iterating the saddle point equation once more, by inserting
$c^{\mbox{\tiny{\sc sda}}}(\vec{\phi})$ in the right hand side of
Eq.~(\ref{iter-sda}). The spectrum in the interval
$[-\lambda_c,\lambda_c]$ is again modified with respect to the ones
found in the {\sc ema} and {\sc sda}. However, it still vanishes at $\pm 
\lambda_c$ and keeps a more or less similar form. 
Outside this band, peaks indexed by $k,
l_1,\dots,l_k$ , with $k$ and $l_i$ integers, appear for values of
$\mu  $ that are solutions to the equation:
\begin{equation}
1=\sum_{i=1}^k \frac{1}{\mu  ^2 - l_i \; \mu  \; \sigma(\mu  )} 
\;\;\;\; \mbox{with} \; |\mu  | \geq \lambda_c
\; .
\label{eq_peaks_tla}
\end{equation}
This corresponds to considering a central site having $k$ neighbours,
each having $l_i+1$ neighbours, the latter being treated as part of
the effective medium. One has thus a more precise description of the
environment of the given site since its first neighbours are treated
exactly. Again, if one considers the limit $|\mu  | \to \infty$, for
which $\sigma(\mu  ) \sim 1/\mu  $, Eq.~(\ref{eq_peaks_tla}) yields 
the roots of the characteristic polynomial of a
finite cluster with the same geometry, see Eq.~(\ref{eq_tltree}).  In this
limit, the part of the graph under consideration is not sensitive to
the rest of the sites.

For the same reasons explained at the {\sc sda} level, there are
minimum values of $k$, $\{l_i\}$, for a solution to
Eq. (\ref{eq_peaks_tla}) to exist. As predicted by the qualitative
argument, there can be localized states with $k$ less than $k_{min}$
of the {\sc sda}, provided the $l_i$s are not too small. For
instance, if $p=10$, there are solutions with $k=20$ and all the
$l_i$s equal to $13$.   

Even if  we have not derived an explicit 
asymptotic expression of the density of
state at this level of approximation, we believe that the qualitative 
features of     $\rho^{\mbox{\tiny{\sc sda}}}(\mu  )$ will not be modified. 
Indeed, large eigenvalues are due to sites having a
connectivity much larger than the mean and these are rare
events. Configurations in which 
the connectivity of the first neighbours also differs 
considerably from the one of the effective medium have even 
lower probability, they should then be
negligible and yield only a small correction. 
Yet, it would be interesting to evaluate it and to check 
if it is consistent with the next terms of the
expansion which, in principle,  
can be computed from the non perturbative method
in~\cite{Robr}.

\subsection{Asymmetric distribution}
\label{nonsym}

In this section we discuss the influence of the asymmetry
parameter $a$ that quantitifies the ferromagnetic or 
anti-ferromagnetic character of the interactions, see Eq.~(\ref{disorder}). 

We have shown
rigorously that as long as the underlying graph is tree-like, the
parameter $a$  has strictly no influence on the spectrum. 
Thus, when $p<1$,  in the thermodynamic limit,
$\rho(\mu  )$ is strictly independent on $a$.

When $p>1$, one can show that $a$ has no effect on the 
iterative scheme of resolution of
the saddle point equation. 
Indeed, for $a\ne 1/2$, the effective Hamiltonians becomes:
\begin{equation}
- H_{\mbox{\tiny{\sc eff}}}(c(\vec \phi)) = -\frac{p}{2} + \frac{p}{2} 
\int d\vec \phi \: d \vec \psi \: c(\vec \phi) \: c(\vec \psi) 
\left( a\: e^{-i \vec \phi \cdot    \vec \psi} + (1-a)\: e^{i \vec \phi
\cdot    \vec \psi} \right)
\; .
\end{equation}
Since $c^{\mbox{\tiny{\sc ema}}}(\vec \phi)$ is even with respect to
$\vec \phi \to 
-\vec \phi$, Eq.~(\ref{cubic-eq}) on $\sigma(\mu  )$ is not modified.
Then $c^{\mbox{\tiny{\sc sda}}}(\vec \phi)$ is independent on $a$ and
even so, at all orders of the iteration, the order parameter, and thus
the spectrum, do not depend on $a$.

The independence of $\rho(\mu  )$ on $a$ 
was conjectured in~\cite{Rode} and shown rigorously
in~\cite{Bauer}: the
moments of the spectrum do not depend on $a$, and their growth is
sufficiently slow to determine the density of states.
However, we observed numerically a strong dependence on $a$ in finite 
size matrices (see Section~\ref{numerics}). As practical applications of 
random matrix theory are often confined to small matrix sizes, this
cannot be neglected.

\section{Numerical results}
\label{numerics}

In this Section we present the spectra obtained from the 
exact numerical diagonalisation of finite size random matrices. 

\begin{figure}[h]
\centerline{
\input{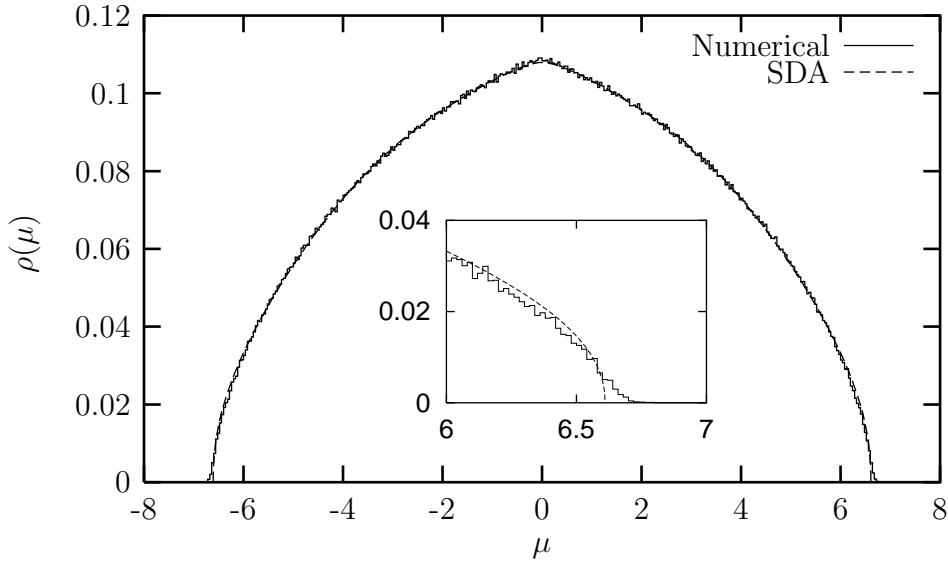}}
\caption{Density of states averaged over
$500$ samples for $p=10, a=1/2$, the matrix size is
$N=2000$. The dashed curve represents the {\sc sda} result for this
value of $p$ and it is almost totally superposed to the numerical data
apart from a small deviation near the edges, shown in the inset.}
\label{figure_rho_numeric}
\end{figure}

In Fig~\ref{figure_rho_numeric} we plot the average density of states,
computed on $500$ samples of $2000 \times  2000$ matrices, with mean
connectivity $p=10$ and a symmetric distribution of the matrix elements,
$a=1/2$. The agreement with the {\sc sda} prediction is very good in
most of the spectrum. However, there is a tail extending beyond
$\lambda_c$, shown in the inset. 
We have checked that this tail was not due to finite-size effects by
repeating the diagonalization for different values of $N$. We have
also checked that eigenvalues in the tail are due to localized states,
by computing the inverse participation ratio for these states, which
does not scale as $1/N$ in the large $N$ limit. However, the height of the
peaks predicted within the {\sc sda} is much too small to explain this tail
quantitatively. We believe that successive iterations of the
approximation scheme are necessary to describe it, by taking into
account the ``cooperative effect'' of neighbours of a site of large
connectivity having connectivities slightly higher than the mean. 
As argued in Section~\ref{replicas}  we 
expect the {\sc sda} to be correct at large eigenvalues. However, the
weight in the asymptotic tail is so small that it would require very
large matrix sizes and a very large number of samples to be
observed. Because of the high computational cost of matrix
diagonalization, it seems impossible to study this phenomenon
directly. In the conclusion we propose  an alternative method to analyse
it. 

\begin{figure}[h]
\centerline{
\input{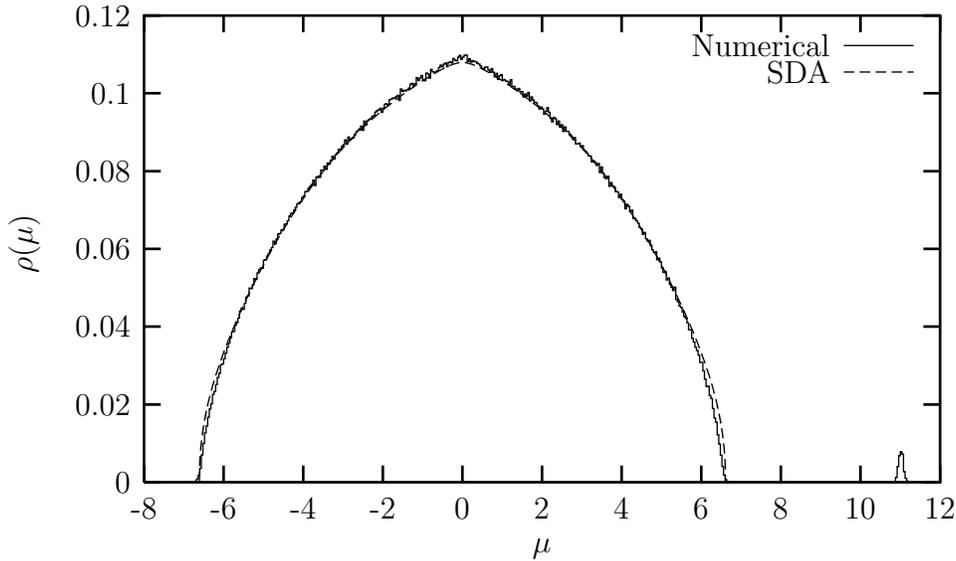}}
\caption{Density of states averaged over
$2000$ samples 
for $p=10, a=1$ and a matrix size $N=800$, with a dashed curve the 
{\sc sda} result for this value of $p$. Note the ferromagnetic bump at
$\mu  \approx 11$.}
\label{figure_rho_num_ferro}
\end{figure}

We have also studied numerically the influence of the asymmetry
parameter $a$ on the average density of states. In
Fig.~\ref{figure_rho_num_ferro} we plot the average density of states,
computed on $2000$ samples of $800 \times  800$ matrices, with mean
connectivity $p=10$ and a purely ferromagnetic distribution, $a=1$.  
We find a very similar spectrum in the region
$[-\lambda_c,\lambda_c]$ with the addition of a separated bump at larger
eigenvalues. The total weight in the bump is, up to numerical
precision, $1/N$ (we have checked this scaling for different values of
$N$), suggesting that for each realization of the matrix, the largest
eigenvalue is in the bump. This assumption was confirmed directly,
by treating separately the largest eigenvalue of each sample. 

For $N=800$ and $p=10$ we found that the location of the largest eigenvalue
decreases smoothly from $a=1$ to $a \approx 0.65$,
where it approximately approaches the edge of the $a=0.5$ 
distribution. This observation is not contradictory with the 
previous statements on
the independence on $a$ of the density of states in the thermodynamic limit,
as this effect is of order $1/N$. 
A detailed study of the $N$ and $a$ dependence
is delicate and it goes beyond the scope of this work. 
Yet, it is important to keep in mind this effect
for realistic implementations
where finite size effects are important.

\section{Conclusions}
\label{conclusions}

The study of random matrices is a very rich field. For sparse
matrices, there has been interest in different ensembles
(Laplacians~\cite{Robr2,Bimo,INM}, incidence matrices~\cite{Robr},
``real-world'' graphs~\cite{Bara}, banded matrices~\cite{band}), and in
different properties (large eigenvalue tails, behaviour around small
eigenvalues, quantum/classical percolation~\cite{Evec}). 
In this paper we focused on the specific problem of the 
large eigenvalue tails. We studied the properties of finite-size
clusters and derived recursion relations on characteristic
polynomials. This allows one to solve exactly the problem for low
concentrations $p$.
In addition we used a replica variational method~\cite{Bimo} to study
the large $p$ limit. The combination of these two methods gives
a very simple geometric content to the asymptotic form of the
eigenvalue tails previously obtained with involved
calculus~\cite{Robr}. It is worth mentioning though a drawback of the
method we used: it is not straightforward to estimate the range of
validity of the result and to bound the corrections that successive
levels of iterations generate.

As mentioned in Section~\ref{numerics}, 
the direct exploration of the tails in the
eigenvalue spectrum by means of numerical diagonalisation is a very
difficult task, as these tails are due to very rare events. 
In a forthcoming paper~\cite{Secu} 
we shall use an indirect method that might make
this study easier. Namely, we shall consider the dynamics
of a spherical model with finite connectivity
interactions, whose dynamics is directly related to the eigenvalue
spectrum considered here. 
The dynamics in the long time limit is 
controled by the large eigenvalue behaviour of the density of states and 
its study will hence give us information about the tails of the 
spectrum itself.

\vspace{1cm}
\noindent\underline{Acknowledgments}
 
Very useful discussions with Giulio Biroli and R{\'e}mi Monasson 
are gratefully acknowledged.
GS and LFC acknowledge financial support from the 
ACI ``Algorithmes d'optimisation et syst{\`e}mes d{\'e}sordonn{\'e}s
quantiques''.
LFC acknowledges financial support from the research contract between 
the CNRS (France) and CONICET (Argentina), thanks F. A. Schaposnik and 
the Physics Department at the Universidad Nacional de La Plata, Argentina,
for hospitality during the last part of this work and the Guggenheim
Foundation. LFC is associate researcher at ICTP, Trieste, Italy.

\end{document}